\documentclass[12pt]{iopart}

\usepackage{graphicx}
\usepackage{bm}        
\usepackage{amssymb}   

\usepackage{iopams}  
\begin{document}

\title[]{Eigenvalues and Eigenfunctions of Two Coupled Normal Metal Nano-rings}

\author{L Fang$^{1,2}$ and D Schmeltzer$^{1,2}$}

\address{$^1$Physics Department, The City College of CUNY, New York, NY 10031, USA}
\address{$^2$The Graduate Center, CUNY, 365 Fifth Avenue, New York, NY 10016, USA}
\ead{lfang@ccny.cuny.edu}
\vspace{10pt}
\begin{indented}
\item[]July 2015
\end{indented}

\begin{abstract}
A general scheme is developed to deal with 1D lattice systems that could be topologically complicated. It is aimed to give a complete study of two coupled normal metal rings. Our method starts with an investigation of the local expressions of the eigenfunctions. By connecting different parts of the system, all the eigenvalues and eigenfunctions can be obtained. It is found that there is a possibility for the existence of localized states, which is beyond previous expectations.

\vspace{1pc}
\noindent  Keywords: {\it two coupled rings, transfer matrix, localized state}
\end{abstract}
\pacs{03.65.-w,76.23.Ra,73.21.-b}
%

%
%
%

\section{Introduction}

The tight binding model is widely used to study electron behaviours in crystals. While it is originally suited for real materials that are of 3D, people are tempted to study the 1D case for simplicity. Usually only bulk properties are considered, so the Born-von Karman periodic boundary condition is chosen for convenience. The dispersion $\epsilon=-2t\cos k$ is well-known for this 1D periodic tight-binding model \cite{AshcroftMermin}. Although open boundary conditions are most encountered in reality, the argument often used is that the physics in the bulk should not depend on any boundary conditions within the thermodynamic limit. 

As technology develops, ever smaller systems can be manufactured, and realistic 1D systems like quantum wires are realized. It happens that certain systems become sensitive to their boundary conditions due to small sizes. Additionally, the 1D models that theorists could only play with previously come to be relevant. In this background, some combined 1D systems have been proposed, for example two coupled rings \cite{DavidCoupledRings}\cite{Avishai}. To deal with models with complicated boundary conditions, a systematic method is in need. 

In previous works on the two coupled rings, several different methods have been used. In \cite{DavidCoupledRings}, the method of Dirac constraints\cite{Dirac} was used. In \cite{Avishai}, a simple wave function ansartz method was used . Since the focus of these papers was on persistent current \cite{AB_phase} \cite{CNYang} \cite{Persistent_Current_in_Normal_Metal_Rings}, the solutions of the coupled system were discussed only briefly and not completely. The present work is dedicated to outline a general scheme to solve complicated 1D systems. By ``solve", it is meant to obtain all the energy levels and the corresponding eigenfunctions. The two coupled rings problem will be solved completely under this scheme. 

Instead of the continuous model which is used in \cite{DavidCoupledRings} \cite{Avishai}, a lattice model is applied in this work. The physics has no essential difference between these two models, but the latter one has two advantages. First, the lattice model follows exactly the spirit of tight binding, such that when the system becomes small to the extent of several nanometers, the discreteness of atoms may become important. Second, the results could be checked easily by diagonalizing Hamiltonian matrices numerically.  

The usual way of doing quantum mechanics in 1D is to divide the system into several homogeneous parts, obtain the general form of the wave function for each part first, and then match them at the boundaries \cite{LandauQM}.  We will employ this standard procedure to study two coupled rings. Before we jump into this complicated problem, it is pedagogical and beneficial to start from simple situations with simple boundary conditions. Thus we would like to give a systematic review of 1D tight binding models, and see how different boundary conditions could be treated.

We will begin with the semi-infinite wire in Section 2. Here the single open end of the wire is the natural starting point for us to obtain the eigenfunction by iteratively applying the Schroedinger equation. We will use a transfer matrix method to do this iteration. The general form of the eigenfunction is determined by calculating the power of the transfer matrix. Since there is only one boundary condition for the semi-infinite wire, the eigenfunction is obtained readily. After this model is fully understood, we will consider more complicated boundary conditions. 

In Section 3 we study a finite open wire with two open ends. We use the same scheme by starting from one end and using the transfer matrix to obtain a general form of the eigenfunction, but now the eigenfunction should also fulfil the boundary condition on the other open end. This gives us an additional equation which gives a constraint to the eigenvalues and so determines a discrete spectrum. In contrast, the semi-infinite wire has no such equation and it has a continuous spectrum.

In Section 4 we come to study a single closed ring where the boundary condition is periodic. The result for this case is well known, since a Fourier transform can be applied to solve it very quickly. While now we don't want to use the Fourier transform method, we want to perform the same procedure as we have done for the finite open wire. This offers us an opportunity to test our scheme by comparing our results to the familiar ones. The boundary condition now is distinct from the finite open wire, so we have to obtain a different equation that determines the eigenvalues. An interesting point for the single closed ring is that we can add magnetic flux through it, which leads to the famous Aharonov-Bohm effect \cite{AB_phase} and persistent currents \cite{Persistent_Current_in_Normal_Metal_Rings}. For this situation, a gauge transform can be used to give the general result. 


In section 5 we will begin to study two coupled rings. Here we assume the rings are connected by allowing electrons to hop from a site on one ring to a site on the other ring. It is different from the model considered in \cite{DavidCoupledRings} \cite{Avishai}, in which there is a common site for both rings. Actually our current scheme is capable to deal with both situations. But since our goal is only to outline a general scheme and it is not aimed to study all the possible models, we do not consider the common-site model in this work. To solve the problem of the two coupled rings, first we get the local form of the eigenfunction for each single ring, just as what we have done for the single closed ring. Then we use the boundary conditions to connect the local wave functions, where the connection equations are unambiguously written down  by observing the Hamiltonian matrix directly. Similar to the finite open wire and the single closed ring, we obtain an equation that determines the complete spectrum of the system. By studying this equation carefully, we find there may exist localized states. The possibility of this fact is ignored in previous works \cite{DavidCoupledRings} \cite{Avishai}.



\section{Semi-Infinite Wire}

We use the single band tight-binding model throughout this paper. It means on each lattice site there is one and only one state for a single electron. An electron can hop from one site to its nearest neighbours. Electrons are assumed spinless and non-interacting. The many particle states are composed by occupying single electron levels one by one.   

\begin{figure}
\begin{center}
\includegraphics[scale=0.5]{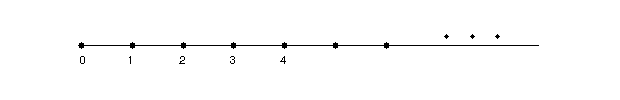}
\end{center}
\caption{ Lattice model for the semi-infinite open wire. The solid spots on the line represent lattice sites. They are linked by allowing electrons to hop between nearest neighbour sites.} \label{fig:SemiInfiniteWire}
\end{figure}

Our starting point is a semi-infinite wire, of which the configuration is shown in Figure \ref{fig:SemiInfiniteWire}. The most notable feature in this figure is that there is a single open end for the semi-infinite wire. The Hamiltonian under our assumption can be written in the formalism of second quantization as  
\begin{equation}
\hat{\mathcal{H}} = -t\sum_{i=0}^{\infty}\hat{a}^{\dagger}_{i+1}\hat{a}_{i} + \mathrm{h.c.} 
\end{equation}
where $t$ is the hopping constant, $\hat{a}_i^{\dagger}$ and $\hat{a}_i$ are creation and annihilation operators for electrons on site $i$.

Set $\hat{A}=(\hat{a}_0,\hat{a}_1,...)^\mathrm{T}$, and rewrite the Hamiltonian in a matrix form 
\begin{equation} \label{eqn:H_matrix_form_Semi_Wire}
\hat{\mathcal{H}} = -t \hat{A}^{\dagger} \mathit{h} \hat{A}
\end{equation}
where 
\begin{equation}
\mathit{h} = \left( \begin{array}{ccccc}
   0   &    1   &        &        &      \\
   1   &    0   &    1   &        &      \\
       &    1   &    0   &    1   &      \\
       &        &    1   &    0   &      \\
       &        &        &        &\ddots
\end{array} \right)
\end{equation}
is equivalent to the Hamiltonian matrix in the formalism of first quantization. Assume $\psi=(x_0,x_1,...)^\mathrm{T}$ to be an eigenvector of $h$ with eigenvalue $\lambda$, then $[x_0\hat{a}^\dagger_0 + x_1\hat{a}^\dagger_1 + x_2\hat{a}^\dagger_2+\cdots]|\Omega\rangle$ ($|\Omega\rangle$ represents the vacuum state) is a single particle eigenstate of $\hat{H}$ with energy level $\epsilon=-\lambda t$. We would like to diagonalize $h$, that is to say, we want to find all the eigenvalues of $h$ and the corresponding eigenvectors. In this way we can obtain all the single particle eigenstates of $\hat{H}$. 

The eigenvalue equation of $h$ reads
\begin{equation} \label{eqn:Semi_Wire_Eigen_Eq}
\left( \begin{array}{ccccc}
   0   &    1   &        &        &      \\
   1   &    0   &    1   &        &      \\
       &    1   &    0   &    1   &      \\
       &        &    1   &    0   &      \\
       &        &        &        &\ddots
\end{array} \right)
\left(
\begin{array}{c}
x_0\\x_1\\x_2\\x_3\\ \vdots 
\end{array}
\right)
  = \lambda
\left(  
\begin{array}{c}
x_0\\x_1\\x_2\\x_3\\ \vdots 
\end{array}
\right)     
\end{equation}
It is also the time independent Schroedinger equation in the formalism of first quantization.
Write equation (\ref{eqn:Semi_Wire_Eigen_Eq}) explicitly 
\begin{eqnarray}
x_1 = \lambda x_0               \nonumber  \\
x_0+x_2 = \lambda x_1           \nonumber  \\
x_1+x_3 = \lambda x_2           \nonumber  \\
x_2+x_4 = \lambda x_3           \nonumber  \\
\cdots \cdots \cdots            \nonumber  
\end{eqnarray}
Now if we set $x_0=1$, then $x_1=\lambda$. For $n>1$, $x_n=\lambda x_{n-1} - x_{n-2}$. By combining an obvious identity $x_{n-1}=x_{n-1}$, we have
\begin{equation} \label{eqn:iteration}
\left(
\begin{array}{c}
x_n \\ x_{n-1}
\end{array}
\right) 
= \mathit{L}
\left(
\begin{array}{c}
x_{n-1} \\ x_{n-2}
\end{array}
\right)
\end{equation} 
where
\begin{equation} \label{eqn:L}
\mathit{L} = \left(
\begin{array}{cc}
\lambda & -1 \\ 
    1   &  0 
\end{array} \right)
\end{equation}
is the so-called transfer matrix.

$L$ is a constant matrix, independent of $n$. We can iterate equation (\ref{eqn:iteration}) and express $(x_n,x_{n-1})^\mathrm{T}$ in terms of $(x_1,x_0)^\mathrm{T}$ as
\begin{equation} \label{eqn:Iteration}
\left(
\begin{array}{c}
x_n \\ x_{n-1}
\end{array}
\right) 
= \mathit{L}^n
\left(
\begin{array}{c}
x_{1} \\ x_{0}
\end{array}
\right)
\end{equation}

Now if we know $\mathit{L}^n$ we can obtain the entire wave function. The calculation of $L^n$ depends on the value of $\lambda$. We summarize the results here and put its derivation in Appendix A.

\vspace{10pt}
\noindent
i) if $\lambda \in (-2,2)$, set $\lambda = 2\cos\theta$ , where $\theta \in (0,\pi)$, then
\begin{equation} \label{eqn:recursion1}
\mathit{L}^n = \left(
\begin{array}{cc}
\frac{\sin(n+1)\theta}{\sin\theta} & -\frac{\sin n\theta}{\sin\theta} \\
\frac{\sin n\theta}{\sin\theta}    & -\frac{\sin (n-1)\theta}{\sin\theta}
\end{array} \right)
\end{equation}
\vspace{10pt}
ii) if $\lambda=2$, then
\begin{equation} \label{eqn:recursion2}
\mathit{L}^n = \left(
\begin{array}{cc}
n+1 & -n \\ n & -(n-1)
\end{array} \right)
\end{equation}
which corresponds to $\theta\rightarrow 0$ in (\ref{eqn:recursion1}).
\\[5pt]
if $\lambda=-2$, then
\begin{equation} \label{eqn:recursion3}
\mathit{L}^n = (-1)^n
\left( 
\begin{array}{cc}
n+1 & n \\ -n & -(n-1)
\end{array}
\right)
\end{equation}
which corresponds to $\theta\rightarrow\pi$ in (\ref{eqn:recursion1}).
\\[10pt]
iii) if $\lambda>2$, set $\lambda=2\cosh k$, where $k>0$, then
\begin{equation} \label{eqn:recursion4}
\mathit{L}^n = \left(
\begin{array}{cc}
\frac{\sinh(n+1)k}{\sinh k} & -\frac{\sinh nk}{\sinh k} \\
\frac{\sinh nk}{\sinh k}    & -\frac{\sinh (n-1)k}{\sinh k}
\end{array} \right)
\end{equation}
which corresponds to (\ref{eqn:recursion1}) if we set $\theta=\rmi k$
\\[5pt]
if $\lambda<-2$, set $\lambda=-2\cosh k$, where $k>0$, then
\begin{equation} \label{eqn:recursion5}
\mathit{L}^n = (-1)^n
\left(
\begin{array}{cc}
\frac{\sinh(n+1)k}{\sinh k} & \frac{\sinh nk}{\sinh k} \\
-\frac{\sinh nk}{\sinh k}     & -\frac{\sinh (n-1)k}{\sinh k}
\end{array} \right)
\end{equation}
which corresponds to (\ref{eqn:recursion1}) if we set $\theta=\pi+\rmi k$
\\[5pt]

Once we know $L^n$, we can plug $L^n$ into (\ref{eqn:Iteration}) and use the initial value $(x_0,x_1)=(1,\lambda)$ to write down the eigenfunction for each case as above:
\\[5pt]
i) if $\lambda = 2\cos\theta \in (-2,2)$ , $\theta \in (0,\pi)$
\begin{equation}
\psi = 
(1 , 2\cos\theta , \frac{\sin 3\theta}{\sin\theta} , \frac{\sin 4\theta}{\sin\theta} , \cdots)^\mathrm{T}
\end{equation}
Note that $U_n(\cos\theta)\equiv\frac{\sin(n+1)\theta}{\sin\theta}$ is the second kind of Chebyshev polynomial \cite{Chebyshev}.  
\\[10pt]
ii) if $\lambda=2$,
\begin{equation}
\psi = (1 , 2 , 3 , 4 , \cdots)^\mathrm{T}
\end{equation}
if $\lambda=-2$,
\begin{equation}
\psi = 
(1 , -2 , 3 , -4 , \cdots)^\mathrm{T}
\end{equation}
\vspace{10pt}
iii) if $\lambda>2,\lambda=2\cosh k, k>0$
\begin{equation}
\psi = 
(1 , 2\cosh k , \frac{\sinh 3k}{\sinh k} , \frac{\sinh 4k}{\sinh k} , \cdots)^\mathrm{T}
\end{equation}
\vspace{5pt}
if $\lambda<-2,\lambda=-2\cosh k, k>0$
\begin{equation}
\psi = 
(1 , -2\cosh k , \frac{\sinh 3k}{\sinh k} , -\frac{\sinh 4k}{\sinh k} , \cdots)^\mathrm{T}
\end{equation}
\vspace{5pt}

In quantum mechanics wave functions are required to be normalizable, so they cannot blow up at infinity. Solutions ii) and iii) therefore should be abandoned. Thus the spectrum for a semi-infinite wire can only be in the range of $(-2,2)$. Since we have no other constraint to the eigenvalue now, any value in $(-2,2)$ belongs to the spectrum, so the spectrum of the semi-infinite wire is continuous.

In i), if we set $x_0=\sin\theta$, the eigenfunction becomes
\begin{equation}
\psi = 
(\sin\theta , \sin 2\theta , \sin 3\theta , \sin 4\theta , \cdots)^\mathrm{T}
\end{equation}
This form of the eigenfunction looks like a standing wave with only one fixed end. It can be considered as the superposition of the initial wave and the reflected wave when an electron travels from infinity to the boundary and then reflects back.   

\vspace{6pt}
Now we have finished the discussion of the semi-infinite wire. The key to the solution in this section is that we can express the entire wave function in terms of $(x_0,x_1)^\mathrm{T}$ through (\ref{eqn:Iteration}). We note that the simple geometry of the semi-infinite wire gives us two great advantages. First, the single end of the wire supplies a natural starting point to begin the iteration. (Although in principle we can start from any two adjacent sites $(x_n,x_{n+1})^\mathrm{T}$ and derive the entire wave function, it makes calculation more complicated.) Second, once we have solved for the eigenfunction, we need only further to consider the requirement of normalization, since there are no other boundary conditions. 

In the following we want to vary the boundary conditions little by little, and see how our scheme can be adjusted readily for new situations. 

\section{Finite Open Wire}

\begin{figure} [h]
\begin{center}
\includegraphics[scale=0.5]{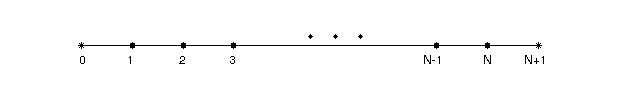}
\end{center}
\caption{ Lattice model for the finite open wire. The solid spots in the middle of the line represent real sites and the star-like spots at the two ends represent imaginary sites.} \label{fig:OpenWire}
\end{figure}

In this section we discuss the finite open wire, the configuration of which is shown in Figure \ref{fig:OpenWire}. Compared to the semi-infinite wire, it has two open ends and only a finite number of sites. Assume the number of sites is $N$, and denote the wave function as $\psi=(x_1,...,x_N)^\mathrm{T}$. 

We would like to follow the same steps as before and start from the open end, since there is no difference in the local region of the end as compared to the semi-infinite wire. Let's begin with $x_1$ and $x_2$ and use the transfer matrix to get the value of $x_3$, $x_4$, $\cdots$, $x_N$. 
Now the difference from the semi-infinite wire is that we cannot go any further, because no site $(N+1)$ exists. We have to stop here and require $x_{N-1}=\lambda x_N$, which is the boundary condition at the other end of the wire. 

It's appropriate to use a little trick at this point. Imagine that we add site $0$ at the left end and site $(N+1)$ at the right end, and require that $x_0 = x_{N+1}=0$. Then the boundary conditions $x_2=\lambda x_1$ and $x_{N-1}=\lambda x_N$ could be tailored as $x_2=\lambda x_1-x_0$ and $x_{N+1}=\lambda x_N-x_{N-1}$, which fits the general form of the iteration equation $x_{n+1}=\lambda x_n-x_{n-1}$. The advantage of this trick is that we have two much simpler boundary conditions $x_0 = x_{N+1}=0$ now.
  

We start from $x_0$( $=0$) and $x_1$, use the transfer matrix and get a general expression of $x_n$, for any $n$. The form of $x_n$ should be the same as that for the semi-infinite wire. Actually we can set $x_1=1$, so $x_2=\lambda$, and then the same iterations as that of the semi-infinite wire follow. 

After obtaining the expression $x_n$ we need to require $x_{N+1}=0$. It's obvious that there is no means for $x_n$ to increase as $n$ increases, so we have to abandon solutions ii) and iii). The eigenvalue $\lambda$ stays in the range $(-2,2)$ just like the case of the semi-infinite wire. For the semi-infinite wire any numbers in $(-2,2)$ are possible spectrum, but now the additional equation $x_{N+1}=0$ is a constraint to $\lambda$ and selects some specific values in $(-2,2)$. 

Set $\lambda=2\cos\theta$ ($0<\theta <\pi$), and $x_0=0$, $x_1=\sin\theta$. Applying the transfer matrix (\ref{eqn:recursion1}), we have $x_n=\sin n\theta$. From $x_{N+1}=0$, there is $\sin (N+1)\theta=0$. This equation has $N$ distinct roots
\begin{equation}
\lambda_m  = 2\cos\theta_m, \hspace{9pt} \theta_m = \frac{m}{N+1}\pi, \hspace{9pt} m=1,2,...,N
\end{equation}
They correspond to energy levels $\epsilon_m = -\lambda_m t$ which constitute the complete spectrum. The eigenfunction with the energy level $\epsilon_m$ is
\begin{equation}
\psi_m \sim 
(0 , \sin\theta_m , \sin 2\theta_m , \cdots , \sin N\theta_m , 0)^\mathrm{T} ,
\end{equation}
up to a normalization constant. It represents a standing wave that has two fixed ends. 

Notice that the continuous version of our lattice model of the finite open wire is that of the particle in a box. From their solutions, the similarity is obvious.

\section{Single Closed Ring}

\begin{figure} [h]
\begin{center}
\includegraphics[scale=0.5]{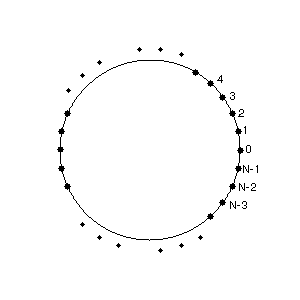}
\caption{ Lattice model for the single closed ring. The solid spots on the circle represent lattice sites.} \label{fig:SingleClosedRing}
\end{center}
\end{figure}

If we connect the two ends of the finite open wire by allowing electrons to hop between the two end sites, then it forms a single closed ring. This configuration, as shown in Figure \ref{fig:SingleClosedRing}, is equivalent to a 1D lattice with the Born-von Karman periodic boundary condition.  
Due to the translational symmetry, it's usually solved through Fourier transform.
Now we want to develop our scheme to deal with the single closed ring and compare the results to the familiar ones. 

Assume there are $N$ sites on the ring, with the wave function $\psi=(x_0,x_1,...,x_{N-1})^T$. Just like what we have done in previous sections, we can start from $x_0$, $x_1$ and use the transfer matrix to attain $x_n$, for arbitrary $n$. Since now we have a periodic boundary condition, we expect that when $n$ is larger than $N$, $x_n$ should be the same as $x_{n-N}$. So we have $x_N=x_0$ and $x_{N+1}=x_1$. For 
\begin{equation}
\left( \begin{array}{c} x_{N+1} \\ x_N \end{array} \right)
= \mathit{L}^N
\left( \begin{array}{c} x_1 \\ x_0 \end{array} \right)
\end{equation}
thus


\begin{equation} \label{eqn:SingleRing_x1_x0}
\left( \begin{array}{c} x_1 \\ x_0 \end{array} \right)
= \mathit{L}^N
\left( \begin{array}{c} x_1 \\ x_0 \end{array} \right)
\end{equation}

\begin{figure}
\begin{center}
\includegraphics[scale=0.45]{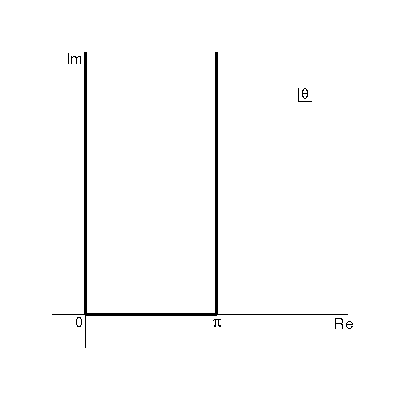} 
\caption{ The domain of $\theta$ in complex plane under the generalized interpretation of $\theta$. The thick U-shape line in the figure represents the domain, which unifies all different situations.}\label{fig:domain}
\end{center}
\end{figure}

\vspace{5pt}
Before we move on, it's useful to unify all three different situations of $\lambda$ by enlarging the domain of $\theta$ in i). As mentioned when we wrote down the expression of $L^n$ in section 2, we can extend the domain of $\theta$ into the complex plane, as shown in Figure \ref{fig:domain}, to include all different situations. In the following we will use this generalized interpretation of $\theta$, and therefore $L^n$ could be simply expressed as (\ref{eqn:recursion1}).

\vspace{5pt}
Now insert (\ref{eqn:recursion1}) into (\ref{eqn:SingleRing_x1_x0}) and we have
\begin{equation} \label{eqn:Ring_x1_x2}
 \left(
\begin{array}{cc}
\frac{\sin(N+1)\theta}{\sin\theta}-1 & -\frac{\sin N\theta}{\sin\theta}       \\
\frac{\sin N\theta}{\sin\theta}      & -\frac{\sin (N-1)\theta}{\sin\theta}-1
\end{array} \right)
\left(
\begin{array}{c}
x_1 \\ x_0
\end{array}\right)
= 0
\end{equation}
If $x_0$ and $x_1$ are both zero, then $x_n=0$ for any $n$. Thus, in order to get a non-zero eigenfunction, we have to first make sure (\ref{eqn:Ring_x1_x2}) has a non-zero solution. So
\begin{equation}
\left| \begin{array}{cc}
\frac{\sin(N+1)\theta}{\sin\theta}-1 & -\frac{\sin N\theta}{\sin\theta}       \\
\frac{\sin N\theta}{\sin\theta}      & -\frac{\sin (N-1)\theta}{\sin\theta}-1
\end{array} \right|
= 0
\end{equation} 
The roots of this equation represent the spectrum of the single closed ring. After fixing $\theta$, the ratio of $x_1$ and $x_0$ could be readily obtained through (\ref{eqn:Ring_x1_x2}). The exact values of $x_1$ and $x_0$, up to a globe phase, can be determined by normalization. The eigenfunction can be expressed in terms of $x_0$ and $x_1$ through (\ref{eqn:recursion1}) as
\begin{equation}\label{eqn:xExpression}
x_n=\frac{\sin n\theta}{\sin\theta}x_1 - \frac{\sin(n-1)\theta}{\sin\theta}x_0
\end{equation}
\vspace{5pt}

We don't give the detailed form of the eigenvalue and the eigenfunction for the single closed ring right now. We want to generalize the model by letting magnetic flux thread the ring. We will study this generalized model carefully. The solution of this generalized model will automatically include that of the ring with no magnetic flux as a special circumstance.

Let's add magnetic flux $\Phi$ through the ring. (This makes the system a setup for the Aharonov-Bohm effect and the persistent current.) The well-known Peierls substitution can be used to include the effect of the magnetic flux into the tight-binding model. Simply speaking, the Peierls substitution varies the hopping constant $t$ by a phase $\phi$, such that $t$ is substituted by $t\rme^{\rmi\phi}$. The phase $\phi$ is related to the vector potential $\mathbf{A}$ via $\phi=(q/\hslash)\int^{r_{i+1}}_{r_i}\mathbf{A} \cdot \rmd \bi{r}$. We can choose a gauge to let $\mathbf{A}$ have the same magnitude along the ring, and so the result does not depend on the site number $i$ in the expression of $\phi$. Now $\Phi = \oint\mathbf{A}\cdot \rmd \bi{r}$ is the total flux. If $\Phi\neq 0$, then $\mathbf{A} \neq 0$ and thus $\phi \neq 0$.     

After the introduction of magnetic flux, the Hamiltonian for the single closed ring can be written as
\begin{equation}
\hat{\mathcal{H}} = -\sum_{i=0}^{N-1} t \rme^{\rmi\phi} \hat{a}^{\dagger}_{i+1} \hat{a}_i + \mathrm{h.c.}
\end{equation} 
where $\hat{a}_N=\hat{a}_0$ is assumed for the ring's topology. 


We could turn this into a matrix form as  (\ref{eqn:H_matrix_form_Semi_Wire})      and would like to diagonalize the first quantization Hamiltonian matrix $h_\phi$. To emphasize the difference, we change the notation a little and set the eigenfunction of $h_\phi$ to be $(y_0,y_1,\cdots,y_{N-1})^\mathrm{T}$.
 
Now the Schroedinger equation (the eigenvalue equation of $h_\phi$) becomes
\begin{equation} \label{eqn:recursion}
y_{n-1}\rme^{-\rmi\phi}+y_{n+1} \rme^{\rmi\phi} = \lambda y_n 
\end{equation}
where we set $\lambda=2\cos\theta$ in the generalized interpretation of $\theta$. This equation differs from our familiar one $x_{n-1}+x_{n+1}=\lambda x_n$ only by a gauge transform. If we set $y_n=\rme^{-\rmi n\phi}x_n$ then they are the same. Thus, based on the result (\ref{eqn:xExpression}), we have 
\begin{equation} \label{eqn:yn_y0y1}
y_n=\frac{\sin n\theta}{\sin\theta} \rme^{-\rmi(n-1)\phi} y_1 - \frac{\sin(n-1)\theta}{\sin\theta} \rme^{-\rmi n\phi} y_0
\end{equation}

Next, the boundary condition could be expressed similar to (\ref{eqn:SingleRing_x1_x0}) as
\begin{eqnarray}
y_0 =\frac{\sin N\theta}{\sin\theta}\rme^{-\rmi(N-1)\phi}y_1 - \frac{\sin (N-1)\theta}{\sin\theta}\rme^{-\rmi N\phi}y_0 \label{eqn:eqny0} \\
y_1 =\frac{\sin(N+1)\theta}{\sin\theta}\rme^{-\rmi N\phi}y_1 - \frac{\sin N\theta}{\sin\theta}\rme^{-\rmi(N+1)\phi}y_0 \label{eqn:eqny1}
\end{eqnarray}

\begin{figure}
\begin{center}$
  \begin{array}{cc}
    \includegraphics[width=.5\textwidth]{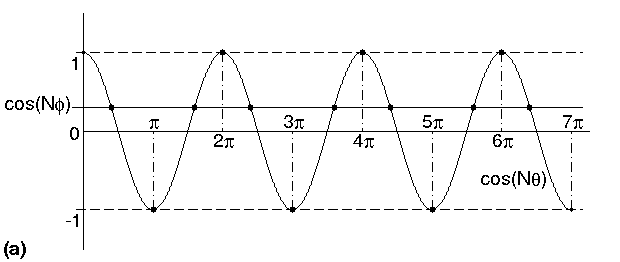}      &
    \includegraphics[width=.5\textwidth]{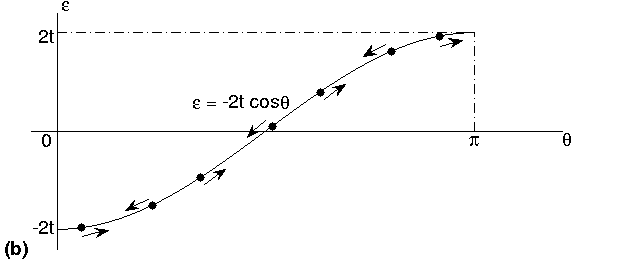}      \\
    \includegraphics[width=.5\textwidth]{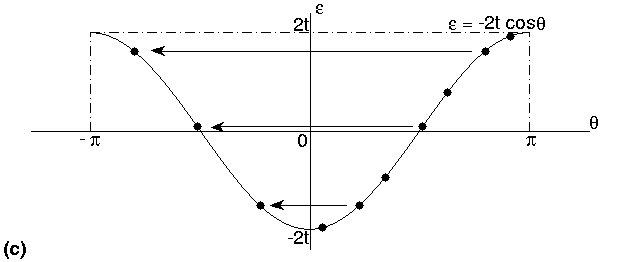}      &
    \includegraphics[width=.5\textwidth]{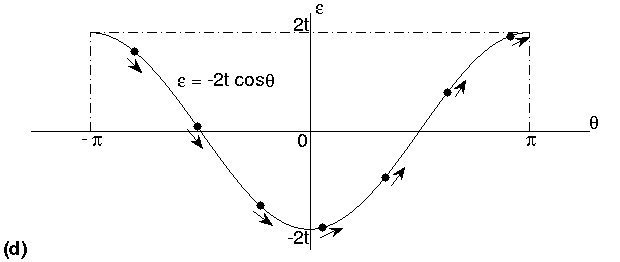}      \\
  \end{array}  $
\end{center}
\caption{ The spectrum of the single closed ring. Note: $N$ is set to be 7 in these plots. (a) Solutions of (\ref{eqn:single_ring_spectrum}). The horizontal axis represents $N\theta$.  (b) Spectrum of the single closed ring. The spots on the curve $\epsilon=-2t\cos\theta$ represent the roots $\frac{m}{N}2\pi\pm\phi$. The arrows around show how the roots move as the magnetic flux increases.     (c)  The roots $(\frac{m}{N}2\pi-\phi)$ are reflected to $-(\frac{m}{N}2\pi-\phi)$ .       (d) The spectrum after the adjustment. The arrows show the direction of the motion under the increase of $\phi$.   } \label{Fig:5}
\end{figure}

\noindent
Although (\ref{eqn:eqny0}) and (\ref{eqn:eqny1}) are equally important now, they are not in the two coupled rings system that we will consider in the next section. Due to the coupling, (\ref{eqn:eqny1}) will be substituted by a different equation. However (\ref{eqn:eqny0}) will still hold for the topology of the closed loop. From (\ref{eqn:eqny0}) we can express $y_1$ in terms of $y_0$
\begin{equation}\label{eqn:y1}
y_1 =  \frac{\rme^{\rmi\frac{N}{2}\phi}\sin\theta + \rme^{-\rmi\frac{N}{2}\phi}\sin(N-1)\theta}{\sin N\theta} \rme^{\rmi(\frac{N}{2}-1)\phi} y_0
\end{equation}
    
Then plug (\ref{eqn:y1}) into (\ref{eqn:yn_y0y1}) , we have
\begin{equation}\label{eqn:yn}
y_{n} = \frac{\rme^{\rmi(N-n)\phi}\sin n\theta + \rme^{-\rmi n\phi}\sin(N-n)\theta}{\sin N\theta} y_0
\end{equation}
One notable property of this wave function is $y_{N-n}=y^*_n$, which means that the system remains the same if we reverse both the direction of the magnetic flux and the direction of the ring. The denominator $\sin N\theta$ may equal to zero, but it occurs only accidentally, corresponding to some degeneracy of the system. It could be lifted by a small perturbation of the flux.

In (\ref{eqn:recursion}) if we set $n=0$ there is
\begin{equation} \label{eqn:SingleRingBC}
y_{N-1}\rme^{-\rmi\phi} + y_1 \rme^{\rmi\phi} = 2\cos\theta y_0
\end{equation}
where the periodic boundary condition is used for the substitution of $y_{-1}$ to $y_{N-1}$. Combine (\ref{eqn:yn}) with (\ref{eqn:SingleRingBC}), after simplification we get 
\begin{equation}\label{eqn:single_ring_spectrum}
\cos N\theta = \cos N\phi
\end{equation}

\noindent
The solution of (\ref{eqn:single_ring_spectrum}) is shown in Figure \ref{Fig:5}(a). Generally we have
\begin{equation}\label{eqn:theta1}
\theta =  \frac{m}{N}2\pi  \pm \phi
\end{equation}
where $m$ belongs to the set of integers resulting in $\theta\in [0,\pi]$. There are total $N$ such integers, which thus constitute the complete spectrum.

Now if $\phi$ varies, the spectrum $\theta$ varies accordingly. But different $\theta$s vary in different directions as shown in Figure \ref{Fig:5}(b), depending on whether they have '$+$' or '$-$' in (\ref{eqn:theta1}). 

We can enlarge the domain of $\theta$ to $[-\pi,\pi]$ and reflect all the $\theta$s with '$-$' sign to the interval $[-\pi,0]$ as shown in Figure \ref{Fig:5}(c)
\begin{equation}
 \frac{m}{N}2\pi-\phi \hspace{5pt} \longrightarrow \hspace{5pt} -(\frac{m}{N}2\pi-\phi) = -\frac{m}{N}2\pi + \phi 
\end{equation}

\noindent
In this way we obtain a unified expression for $\theta$
\begin{equation}\label{eqn:spectrumSingleRing}
\theta = \frac{m}{N}2\pi + \phi
\end{equation}
where $m$ belongs to the set of integers that let $\theta\in [-\pi,\pi]$. Now as $\phi$ varies, all the $\theta$s move in a consistent direction, as shown in Figure \ref{Fig:5}(d).

Plug (\ref{eqn:spectrumSingleRing}) in (\ref{eqn:yn}) and we have
\begin{equation} \label{eqn:wavefunctionSingleRing}
y_n = \rme^{\rmi m(\theta-\phi)} y_0
\end{equation}
To normalize the eigenfunction, we can set $y_0=\frac{1}{\sqrt{N}}$.

Actually (\ref{eqn:spectrumSingleRing}) and (\ref{eqn:wavefunctionSingleRing}) can be easily obtained from a direct Fourier transform. So this justifies our scheme.     

We have finished the discussion of the single closed ring. It serves us a good preparation for the problem of two coupled rings.

\section{Two Coupled Rings}

\begin{figure}
\begin{center}
\includegraphics[scale=0.6]{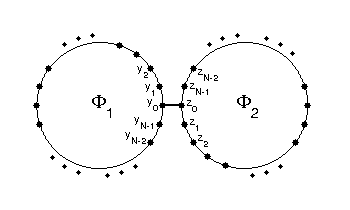}
\caption{Lattice model for the two coupled rings. The solid spots on the circles represent lattice sites. The link in the middle represents the electron can hop between $y_0$ and $z_0$. Magnetic flux $\Phi_1$ is threaded into the left ring and magnetic flux $\Phi_2$ is threaded into the right ring. } \label{fig:TwoRings}
\end{center}
\end{figure}

The configuration of two coupled rings is shown in Figure \ref{fig:TwoRings}. We assume the rings have equal sizes and electrons are allowed to tunnel from the left ring site $y_0$ to the right ring site $z_0$. Magnetic flux $\Phi_1$ is threaded into the left ring and magnetic flux $\Phi_2$ is threaded into the right ring. The Hamiltonian could be cast into
\begin{equation} \label{eqn:H_twoRing}
\hat{\mathcal{H}} = \hat{\mathcal{H}}_1 + \hat{\mathcal{H}}_2 + \hat{\mathcal{H}}_\mathrm{coupling}
\end{equation}
where
\begin{eqnarray}
\hat{\mathcal{H}}_1  =  -\sum_{i=0}^{N-1} [t\rme^{\rmi\phi_1}\hat{a}^{\dagger}_{i+1}\hat{a}_i + \mathrm{h.c.}]  \\
\hat{\mathcal{H}}_2  =  -\sum_{j=0}^{N-1} [t\rme^{\rmi\phi_2}\hat{b}^{\dagger}_{j+1}\hat{b}_j + \mathrm{h.c.}]\\  
\hat{\mathcal{H}}_\mathrm{coupling}  =  -V_0(\hat{a}^{\dagger}_0 \hat{b}_0 + \hat{b}^{\dagger}_0 \hat{a}_0)
\end{eqnarray}
$\hat{\mathcal{H}}_1$ and $\hat{\mathcal{H}}_2$ are two separate single closed ring Hamiltonians. $\hat{\mathcal{H}}_1$ is for the left ring, $\hat{\mathcal{H}}_2$ for the right ring.  $\hat{\mathcal{H}}_\mathrm{coupling}$ represents the coupling between the two rings. In the expressions of $\mathcal{H}_1$ and $\mathcal{H}_2$ above, it is assumed $\hat{a}_N=\hat{a}_0$ and $\hat{b}_N=\hat{b}_0$ for the ring topology.

We emphasize here that the model in \cite{DavidCoupledRings} \cite{Avishai} is different. In \cite{DavidCoupledRings} \cite{Avishai}, $y_0$ and $z_0$ are combined to a single site. This site then connects to $y_1$, $y_{N-1}$, $z_1$ and $z_{N-1}$ separately.

Set the eigenfunction to be $\psi=(y_1,y_2,...,y_{N-1},y_0,z_0,z_1,...,z_{N-1})^\mathrm{T}$. First, we can express the wave function on the left ring in terms of $y_0$ and the wave function on the right ring in terms of $z_0$ separately, exactly the same as what we did for the single closed ring. So like (\ref{eqn:yn}), we could have     
\begin{equation}\label{eqn:twoRing_yn}
y_{n} = \frac{\rme^{\rmi(N-n)\phi_1}\sin n\theta + \rme^{-\rmi n\phi_1}\sin(N-n)\theta}{\sin N\theta} y_0
\end{equation}
\begin{equation}\label{eqn:twoRing_zn}
z_{n} = \frac{\rme^{\rmi(N-n)\phi_2}\sin n\theta + \rme^{-\rmi n\phi_2}\sin(N-n)\theta}{\sin N\theta} z_0
\end{equation}

Next, let's write down the Hamiltonian (\ref{eqn:H_twoRing}) in the matrix form
\begin{equation}
\hat{\mathcal{H}}=-\hat{A}^\dagger h \hat{A}
\end{equation} 
where $\hat{A}=(\hat{a}_1,\hat{a}_2,...,\hat{a}_{N-1},\hat{a}_0,\hat{b}_0,\hat{b}_1,...,\hat{b}_{N-1})^T$, and

\begin{equation} \fl
h = 
\left(
\begin{array}{ccccc|ccccc}
     0       &t\rme^{\rmi\phi_1}&             &             &t\rme^{-\rmi\phi_1}&             &             &                   &                &               \\
t\rme^{-\rmi\phi_1}&            &     ...     &             &             &             &             &                   &                &               \\
             &     ...    &             &t\rme^{\rmi\phi_1} &             &             &             &                   &                &               \\
             &            &t\rme^{-\rmi\phi_1}&     0       &t\rme^{\rmi\phi_1} &             &             &                   &                &               \\
t\rme^{\rmi\phi_1} &            &             &t\rme^{-\rmi\phi_1}&      0      &    V_0      &             &                   &                &               \\ \hline
             &            &             &             &     V_0     &      0      &t\rme^{\rmi\phi_2} &                   &                & t\rme^{-\rmi\phi_2} \\
             &            &             &             &             &t\rme^{-\rmi\phi_2}&     0       &   t\rme^{\rmi\phi_2}       &                &               \\ 
             &            &             &             &             &             &t\rme^{-\rmi\phi_2}&                   &    ...         &               \\ 
             &            &             &             &             &             &             &    ...            &                & t\rme^{\rmi\phi_2}  \\
             &            &             &             &             &t\rme^{\rmi\phi_2} &             &                   &  t\rme^{-\rmi\phi_2} &      0        \\
\end{array}
\right)
\end{equation}
The rows for $y_0$ and $z_0$ show the coupling equations
\begin{eqnarray}
y_{N-1} t e^{-i\phi_1} + y_1 t e^{i\phi_1} + V_0 z_0 = \lambda  y_0 \label{eqn:connection_y0} \\
z_{N-1} t e^{-i\phi_2} + z_1 t e^{i\phi_2} + V_0 y_0 = \lambda  z_0 \label{eqn:connection_z0}
\end{eqnarray}  
where $\lambda$ is the eigenvalue of $h$ (the single level energy is $\epsilon = -\lambda$ under our notation). Set $\lambda=2t\cos\theta$ under the generalized interpretation of $\theta$. Plug (\ref{eqn:twoRing_yn}) and (\ref{eqn:twoRing_zn}) into (\ref{eqn:connection_y0}) and (\ref{eqn:connection_z0}) to get
\begin{equation}\label{eqn:y0z0}
\fl 
\hspace{25pt}
\left( \begin{array}{cc}
2t\sin \theta (\cos N\theta - \cos N\phi_1)  & - V_0\sin N\theta \\
- V_0 \sin N\theta  &  2t\sin \theta (\cos N\theta - \cos N\phi_2)
\end{array}\right)
\left(
\begin{array}{c}
y_0 \\ z_0
\end{array}\right)  =  0
\end{equation}
In order to get a non-zero solution, we have
\begin{equation}
\left| \begin{array}{cc}
2t\sin \theta (\cos N\theta - \cos N\phi_1)  & - V_0\sin N\theta \\
- V_0 \sin N\theta  &  2t\sin \theta (\cos N\theta - \cos N\phi_2)
\end{array}\right|
= 0
\end{equation}
So
\begin{equation} \label{eqn:spectrum_two_rings}
\frac{V_0^2}{4t^2} \frac{\sin^2 N\theta}{\sin^2 \theta} = (\cos N\theta - \cos N\phi_1)(\cos N\theta - \cos N\phi_2)
\end{equation}

\noindent
Equation (\ref{eqn:spectrum_two_rings}) is the main result of this paper. 
The solution of (\ref{eqn:spectrum_two_rings}) determines the spectrum of our two coupled rings system. 

It is noted that $N\phi_1$ is proportional to the total flux that threads into the left ring. Actually $N\phi_1=\frac{\Phi_1}{\Phi_0}\pi$, where $\Phi_0$ is the flux quantum. $N\phi_1$ is the total phase that an electron could acquire due to magnetic flux $\Phi_1$ as it goes around one circle along the left ring. If this phase equals to $2\pi$, there should be no observable effect. Now since the only term involving $\phi_1$ in (\ref{eqn:spectrum_two_rings}) is $\cos N\phi_1$, it's obvious that if we substitute $N\phi_1$ by $N\phi_1+2\pi$, equation (\ref{eqn:spectrum_two_rings}) remains the same. Likewise for $N\phi_2$ and the right ring.

The two rings are coupled through the term $\hat{\mathcal{H}}_\mathrm{coupling}$. If there is no coupling, or to say $V_0=0$ (so $\hat{\mathcal{H}}_\mathrm{coupling}=0$), then we expect to have two separate rings. Now through (\ref{eqn:spectrum_two_rings}) it's clear that if $V_0=0$, we have $\cos N\theta-\cos N\phi_1=0$ or $\cos N\theta - \cos N\phi_2=0$. It means the rings decouple and the electron could only move on one of them.

At this level, we don't try to make a complete analysis of the equation (\ref{eqn:spectrum_two_rings}). It should be very complicated since (\ref{eqn:spectrum_two_rings}) is a transcendental equation. Instead, we want to use an example to show how the solutions of (\ref{eqn:spectrum_two_rings}) look like.

\begin{figure}
\begin{center} $
  \begin{array}{cc}
    \includegraphics[width=.48\textwidth]{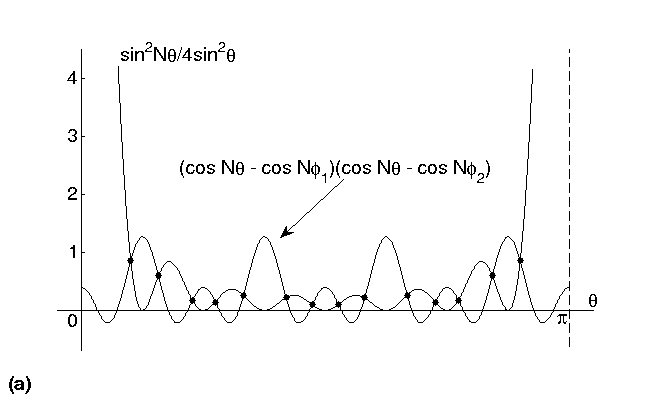} &
    \includegraphics[width=.48\textwidth]{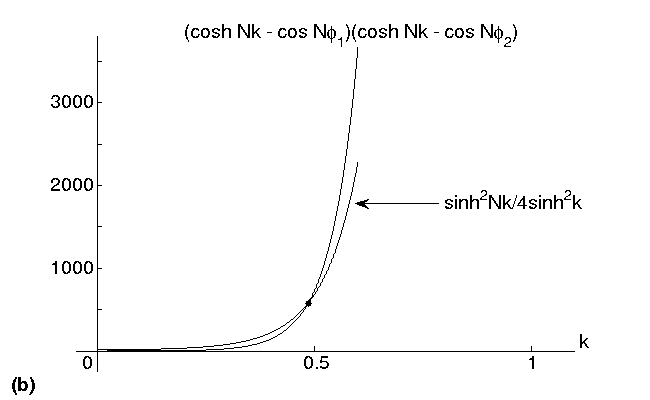}   
  \end{array} $  
\end{center}
  \caption{ Solutions of (\ref{eqn:spectrum_two_rings}), the equation that determine the complete spectrum of the two coupled rings. Note: we set $N=8$, $V_0=t=1$ in this plot. (a) the situation when the eigenvalue $\lambda$ is in the range $[-2,2]$ and we set $\lambda=2\cos\theta$ ; (b) the situation when the eigenvalue $\lambda$ is in the range $[2,\infty)$ and we set $\lambda=2\cosh k$.} \label{fig:TwoRingSpectrum}
\end{figure}

Let $N=8$, $V_0=t=1$ and choose two random fluxes $\phi_1$ and $\phi_2$. 
Figure \ref{fig:TwoRingSpectrum} shows a sketch on how the solutions of (\ref{eqn:spectrum_two_rings}) look with these parameters.

In Figure \ref{fig:TwoRingSpectrum}(a), we show the situation when $\lambda$ is in the range $[-2,2]$ ($\theta$ is real and in the range $[0,\pi]$). The states here are propagating ones, for which the electron could move freely on both rings. As we will see, the wave functions of these states represent charge density waves. It may seem that all the eigenstates should belong to this kind, but that is not the case. There are 16 sites in our setup, so we should expect 16 eigenstates in total. We could find 14 intersections in Figure \ref{fig:TwoRingSpectrum}(a), which correspond to only 14 eigenvalues and eigenstates. Thus there must exist 2 eigenstates that are not propagating states, of which the eigenvalues are beyond the interval $[-2,2]$. 

If $\lambda>2$, we could use the general interpretation of $\theta$ and set it to be $\rmi k$. Figure \ref{fig:TwoRingSpectrum}(b) shows there is a root in this range. So there is an eigenvalue in the range $(2,\infty)$. Similarly, there exists an eigenvalue in the range $(-\infty,-2)$, although we don't show it explicitly here.     

The eigenstates of which the eigenvalues are not in $[-2,2]$ correspond to localized states (bound states). As we will see, in this situation the wave function is localized around the junction and decays exponentially as it goes away from the junction. We have never met such kind of localized states before. It appears neither on the semi-infinite wire, nor on the finite wire, nor on the single closed ring. Also, we note that the possibility of the existence of the localized states has never been mentioned in the previous works on two coupled rings.

\vspace{10pt}

Next we want to discuss the local density of states and the normalization of the wave functions. Now assume $\theta$ is a solution of (\ref{eqn:spectrum_two_rings}).
From (\ref{eqn:twoRing_yn}) and (\ref{eqn:twoRing_zn}), we have
\begin{equation} \label{eqn:twoRing_yDenstiy}
\frac{|y_n|^2}{|y_0|^2} = 
\frac{\sin^2 n\theta + \sin^2 (N-n)\theta + 2\sin n\theta \sin(N-n)\theta \cos N\phi_1}{\sin^2 N\theta}
\end{equation}
\begin{equation} \label{eqn:twoRing_zDenstiy}
\frac{|z_n|^2}{|z_0|^2} = 
\frac{\sin^2 n\theta + \sin^2 (N-n)\theta + 2\sin n\theta \sin(N-n)\theta \cos N\phi_2}{\sin^2 N\theta}
\end{equation}
From (\ref{eqn:y0z0}), there is
\begin{equation} \label{eqn:twoRing_Density_z0/y0}
\frac{|z_0|^2}{|y_0|^2} = \frac{(\cos N\theta - \cos N\phi_1)^2}{(\cos N\theta - \cos N\phi_2)^2}
\end{equation}
Equation (\ref{eqn:twoRing_yDenstiy}), (\ref{eqn:twoRing_zDenstiy}) and (\ref{eqn:twoRing_Density_z0/y0}) show the relative magnitude of the eigenstate's local density. When $\theta$ is in the range $[0,\pi]$, it's obvious that the local density is oscillating on each ring. So it represents a charge density wave and we say that the state is a propagating one. When $\theta$ is out of the range $[0,\pi]$, we could use $k$ to substitute $\theta$ either by $\theta=\rmi k$ (when $\lambda>2$) or by $\theta=\pi+\rmi k$ (when $\lambda<-2$). Then $\cos n\theta$ and $\sin n\theta$ could be substituted by $\pm\cosh nk$ and $\pm\sinh nk$. Since $\cosh nk$ and $\sinh nk$ are basically proportional to $\rme^{nk}$, it's easy to see that the local density is largest at the junction region, and decays exponentially away from it. In this sense, we have a localized state.

We can use the trigonometric identities
\begin{equation*} 
\sum_{n=0}^{N-1} \sin^2 n\theta = \frac{N}{2} - \frac{\sin N\theta \cos(N-1)\theta}{2\sin\theta}
\end{equation*}
\begin{equation*}
\sum_{n=0}^{N-1} \sin^2(N-n)\theta = \frac{N}{2} - \frac{\sin N\theta \cos(N+1)\theta}{2\sin\theta}
\end{equation*}
\begin{equation*}
\sum_{n=0}^{N-1} 2\sin n\theta \sin(N-n)\theta = \cot\theta \sin N\theta - N \cos N\theta
\end{equation*}
to sum (\ref{eqn:twoRing_yDenstiy}) and (\ref{eqn:twoRing_zDenstiy}) for all sites $n$
\begin{equation}\label{eqn:sum_y} \fl \hspace{30pt}
\sum_{n=0}^{N-1} \frac{|y_n|^2}{|y_0|^2}     =    
 \frac{N(1-\cos N\theta \cos N\phi_1) + \cot 
\theta \sin N\theta (\cos N\phi_1 - \cos N\theta)}{\sin^2 N\theta}
\end{equation}
\begin{equation}\label{eqn:sum_z} \fl \hspace{30pt}
\sum_{n=0}^{N-1} \frac{|z_n|^2}{|z_0|^2}     =    
 \frac{N(1-\cos N\theta \cos N\phi_2) + \cot 
\theta \sin N\theta (\cos N\phi_2 - \cos N\theta)}{\sin^2 N\theta}
\end{equation}
We can combine (\ref{eqn:twoRing_Density_z0/y0}), (\ref{eqn:sum_y}) and (\ref{eqn:sum_z}) to normalize the wave function and obtain the values of $|y_0|$ and $|z_0|$. The exact values of $y_0$ and $z_0$ are determined by equation (\ref{eqn:y0z0}) and the normalization of the wave function.  
 Since the final expression is very long and gives little insight, we do not write it explicitly here.


\section{Conclusion}
We have developed a general scheme to study 1D single band tight binding models with complicated boundary conditions. The two coupled rings problem has been analysed step by step in this framework. The eigenvalues and eigenfunctions are given in a complete way. 

Our scheme starts from an investigation of the local property of the wave function. We find that in a homogeneous region, an explicit form of the wave function can be readily obtained in terms of its values on two nearest neighbour sites by a transfer matrix method. Then with the boundary conditions for some special sites expressed specifically, we are able obtain an equation to determine the complete spectrum. 

Note that we divide the range of the eigenvalue $\lambda$ mainly into two different cases. One is in $[-2,2]$, the other is out of $[-2,2]$. For the former, the corresponding eigenstates are propagating ones. For the latter, the wave functions are localized, which represent bound states. We have found that in our model of two coupled rings, there could exist bound states, which are localized around the junction. 


The potential applications of these localized states are unclear at this moment. We expect that if we couple many rings together and form a kind of nano-ring network, then at each junction there exists a localized state. The wave functions of the localized states at different junctions could have small overlaps, and so we expect that electrons are able to tunnel from one junction to a nearby junction. In this way, we have obtained a new lattice, formed by junctions between nearby nano-rings. Obviously, the lattice constant and the hopping constant can be adjusted by changing the size of the nano-rings.   

Finally, we would like to point out that the method we have used in this paper is not restricted to the single band model. It is also possible to be applied to multi-band tight binding models. In such circumstances, the transfer matrices are more complicated. 

\ack

We would like to thank Prof. Klaus Ziegler for mentioning a trick to write down the transfer matrix during one early discussion. We would also like to thank Jeff Sector for editing the draft before it is submitted. 

\appendix
\section{The calculation of $L^n$} 

In this appendix we derive the expression of $L^n$, the result of which is shown in the main text from equation (\ref{eqn:recursion1}) to (\ref{eqn:recursion5}).

\vspace{5pt}
From (\ref{eqn:L}) the transfer matrix $L$ is 
\begin{equation}
\mathit{L} = \left(
\begin{array}{cc}
\lambda & -1 \\ 
    1   &  0 
\end{array} \right)
\end{equation}
where $\lambda$ is a real number, corresponding to the eigenvalue of the original Hamiltonian $h$.

\vspace{10pt} 


Let's try to find the eigenvalue of $L$ first. Assume $(a,b)^\mathrm{T}$ is an eigenvector of $L$ with eigenvalue $\eta$. So

\begin{equation} \label{Eq:eigenEq}
\left(
\begin{array}{cc}
\lambda & -1 \\ 
    1   &  0 
\end{array} \right)
\left(
\begin{array}{c}
a \\ 
b 
\end{array} \right)
= \eta
\left(
\begin{array}{c}
a \\ 
b 
\end{array} \right)
\end{equation} 
or
\begin{equation}
\left(
\begin{array}{cc}
\lambda - \eta & -1      \\
         1     & -\eta
\end{array}
\right)
\left(
\begin{array}{c}
a \\ b
\end{array}
\right) = 0
\end{equation}
In order to get a non-zero eigenvector, we have to require
\begin{equation}
\begin{array}{|cc|}
\lambda - \eta & -1      \\
         1     & -\eta
\end{array}
=0
\end{equation}
or 
\begin{equation}
(\lambda - \eta )(-\eta) + 1 = 0
\end{equation}
Thus
\begin{equation}
\eta = \frac{\lambda \pm \sqrt{\lambda^2 - 4}}{2}
\end{equation}

\vspace{10pt}

Now according to the value of $\lambda$, it is necessary to separate the discussion into three different situations

\vspace{5pt}
\noindent
i) if $-2 < \lambda < 2$

\vspace{5pt}

In this case we can set $\lambda = 2\cos\theta$, where $0 < \theta < \pi$. And so 
\begin{equation}
\eta_{1,2}  = \frac{\lambda \pm \sqrt{\lambda^2 - 4}}{2} = \rme^{\pm \rmi \theta}
\end{equation}

\vspace{3pt}
\noindent
For $\eta_1 = \rme^{i\theta}$, from (\ref{Eq:eigenEq}) the corresponding eigenvector is 
\begin{equation}
\left(
\begin{array}{c}
a_1 \\ b_1
\end{array}
\right)
=
\left(
\begin{array}{c}
\rme^{\rmi\theta} \\ 1
\end{array} 
\right)
\end{equation}

\vspace{3pt}
\noindent
For $\eta_2 = \rme^{-\rmi\theta}$, from (\ref{Eq:eigenEq}) the corresponding eigenvector is 
\begin{equation}
\left(
\begin{array}{c}
a_2 \\ b_2
\end{array}
\right)
=
\left(
\begin{array}{c}
1 \\ \rme^{ \rmi \theta}
\end{array} 
\right)
\end{equation}

\vspace{10pt}

\noindent
Now we can do a liner transform to diagonalize $L$. The transformation matrix is formed as
\begin{equation}
A = \left(
\begin{array}{cc}
a_1 & a_2 \\
b_1 & b_2
\end{array}
\right)
=
\left(
\begin{array}{cc}
\rme^{ \rmi \theta } &      1         \\
    1      & \rme^{ \rmi \theta}
\end{array}
\right)
\end{equation}
and $L$ can be written in the form
\begin{equation}
L = A D A^{-1} =
 A \left(
\begin{array}{cc}
\eta_1 &  \\   &  \eta_2
\end{array}
\right) A^{-1}
\end{equation}
where 
\begin{equation}
D = \left(
\begin{array}{cc}
\eta_1 &  \\   &  \eta_2
\end{array}
\right)= \left(
\begin{array}{cc}
\rme^{\rmi\theta} & \\ & \rme^{-\rmi\theta}
\end{array}
\right)
\end{equation}
and
\begin{equation}
A^{-1} = \frac{1}{\det (A)}
\left(
\begin{array}{cc}
b_2  & -a_2 \\
-b_1 &  a_1 
\end{array}
\right)
= \frac{1}{\rme^{2\rmi\theta}-1}
\left(
\begin{array}{cc}
\rme^{\rmi\theta} &     -1       \\
   -1       & \rme^{\rmi\theta} 
\end{array}
\right)
\end{equation}
is the inverse of $A$.

\vspace{5pt}

Now $L^n$ can be calculated easily
\begin{equation}
L^n = (AD \underbrace{A^{-1})(ADA^{-1})\cdots (A}_n DA^{-1}) = AD^n A^{-1}
\end{equation}
Since
\begin{equation}
D^n = \left(
\begin{array}{cc}
\eta_1^n & \\ & \eta_2^n
\end{array}
\right) =
\left(
\begin{array}{cc}
\rme^{\rmi n\theta} & \\ & \rme^{-\rmi n\theta}
\end{array}
\right)
\end{equation}
We have
\begin{equation}
L^n = \frac{1}{\rme^{2\rmi\theta}-1}
\left(
\begin{array}{cc}
\rme^{ \rmi \theta } &      1         \\
    1      & \rme^{ \rmi \theta}
\end{array}
\right)
\left(
\begin{array}{cc}
\rme^{\rmi n\theta} & \\ & \rme^{-\rmi n\theta}
\end{array}
\right)
\left(
\begin{array}{cc}
\rme^{\rmi\theta} &     -1       \\
   -1       & \rme^{\rmi\theta} 
\end{array}
\right)
\end{equation}
Do this matrix multiplication, and after simplification we obtain
\begin{equation}
L^n = \left(
\begin{array}{cc}
\frac{\sin (n+1)\theta}{\sin \theta} & -\frac{\sin n\theta}{\sin \theta} \\
\frac{\sin n\theta}{\sin \theta}  &  -\frac{\sin(n-1)\theta}{\sin \theta}
\end{array}
\right)
\end{equation}

\vspace{5pt}
\noindent
ii) if $\lambda = \pm 2$

\vspace{5pt}

We only discuss the situation of $\lambda = 2$ here. 
In this case 
\begin{equation}
\eta  = \frac{\lambda \pm \sqrt{\lambda^2 - 4}}{2} = 1
\end{equation}
There is only one eigenvalue. From (\ref{Eq:eigenEq}) we have the corresponding eigenvector
\begin{equation}
\left(
\begin{array}{c}
a \\ b
\end{array}
\right)
=
\left(
\begin{array}{c}
1 \\ 1
\end{array}
\right)
\end{equation}
There is no second linearly independent eigenvector. Thus $L$ cannot be diagonalized. As a compromise, we try to find a vector $(c,d)^T$ that fulfils
\begin{equation}
L \left(
\begin{array}{c}
c \\ d
\end{array}
\right)
= \left(
\begin{array}{c}
a \\ b
\end{array}
\right) + \eta
\left(
\begin{array}{c}
c \\ d
\end{array}
\right)
\end{equation} 
or
\begin{equation}
\left(
\begin{array}{cc}
2 & -1 \\ 1 & 0
\end{array}
\right)
\left(
\begin{array}{c}
c \\ d
\end{array}
\right) =
\left(
\begin{array}{c}
1 \\ 1
\end{array}
\right) + 
\left(
\begin{array}{c}
c \\ d
\end{array}
\right)
\end{equation} 
So
\begin{equation}
\left(
\begin{array}{c}
c \\ d
\end{array}
\right) = \left(
\begin{array}{c}
1 \\ 0
\end{array}
\right)
\end{equation}

\vspace{10pt}
\noindent
Now we form a transformation matrix
\begin{equation}
A = \left(
\begin{array}{cc}
a & c \\ b & d
\end{array}
\right) = \left(
\begin{array}{cc}
1 & 1 \\ 1 & 0
\end{array}
\right)
\end{equation}
and write $L$ in the form
\begin{equation}
L = A D A^{-1} 
= A \left(
\begin{array}{cc}
\eta & 1 \\ 0 & \eta
\end{array}
\right) A^{-1} 
\end{equation}
where
\begin{equation} \label{Eq:D1}
D = \left(
\begin{array}{cc}
\eta & 1 \\ 0 & \eta
\end{array}
\right)
=
\left(
\begin{array}{cc}
1 & 1 \\ 0 & 1
\end{array}
\right)
\end{equation}
and 
\begin{equation}
A^{-1} = \left(
\begin{array}{cc}
0 & 1 \\ 1 & -1
\end{array}
\right)
\end{equation} 
is the inverse of $A$.

\vspace{10pt} 
\noindent
Similar to the case $-2<\lambda<2$, we have
\begin{equation}
L^n = A D^n A^{-1}
\end{equation}
Now
\begin{equation}
D^n = \left(
\begin{array}{cc}
\eta & 1 \\ 0 & \eta
\end{array}
\right)^n = \left(
\begin{array}{cc}
\eta^n & n\eta^{n-1} \\ 0 & \eta^n
\end{array}
\right) = \left(
\begin{array}{cc}
1 & n \\ 0 & 1
\end{array}
\right)
\end{equation}  
So
\begin{equation}
L^n = \left(
\begin{array}{cc}
1 & 1 \\ 1 & 0
\end{array}
\right) \left(
\begin{array}{cc}
1 & n \\ 0 & 1
\end{array}
\right)  \left(
\begin{array}{cc}
0 & 1 \\ 1 & -1
\end{array}
\right) = \left(
\begin{array}{cc}
n+1 & -n \\ n & -(n-1)
\end{array}
\right)
\end{equation}

\vspace{5pt}
\noindent
iii) if $\lambda > 2$ or $\lambda < -2$

\vspace{5pt}

We only discuss the situation of $\lambda > 2$ here.

\vspace{5pt}
\noindent
Set $\lambda = 2 \cosh k$, where $k>0$. So 
\begin{equation}
\eta_{1,2}  = \frac{\lambda \pm \sqrt{\lambda^2 - 4}}{2} = \rme^{\pm k}
\end{equation}
 
\vspace{3pt}
\noindent
For $\eta_1 = \rme^k$, from (\ref{Eq:eigenEq}) the corresponding eigenvector is 
\begin{equation}
\left(
\begin{array}{c}
a_1 \\ b_1
\end{array}
\right)
=
\left(
\begin{array}{c}
\rme^k \\ 1
\end{array} 
\right)
\end{equation} 
 
\vspace{3pt}
\noindent
For $\eta_2 = \rme^{-k}$, from (\ref{Eq:eigenEq}) the corresponding eigenvector is 
\begin{equation}
\left(
\begin{array}{c}
a_2 \\ b_2
\end{array}
\right)
=
\left(
\begin{array}{c}
1 \\ \rme^k
\end{array} 
\right)
\end{equation} 
 
\vspace{10pt}

\noindent
Now we use a liner transform to diagonalize $L$.  Form the transformation matrix
\begin{equation}
A = \left(
\begin{array}{cc}
a_1 & a_2 \\
b_1 & b_2
\end{array}
\right)
=
\left(
\begin{array}{cc}
\rme^k &    1         \\
 1  &   \rme^k
\end{array}
\right)
\end{equation}
and $L$ can be written in the form
\begin{equation}
L = A D A^{-1} =
 A \left(
\begin{array}{cc}
\eta_1 &  \\   &  \eta_2
\end{array}
\right) A^{-1}
\end{equation}
where 
\begin{equation}
D =\left(
\begin{array}{cc}
\eta_1 &  \\   &  \eta_2
\end{array}
\right)= \left(
\begin{array}{cc}
\rme^k & \\ & \rme^{-k}
\end{array}
\right)
\end{equation}
and
\begin{equation}
A^{-1} = \frac{1}{\det (A)}
\left(
\begin{array}{cc}
b_2  & -a_2 \\
-b_1 &  a_1 
\end{array}
\right)
= \frac{1}{\rme^{2k}-1}
\left(
\begin{array}{cc}
\rme^k  &  -1       \\
 -1  &  \rme^k 
\end{array}
\right)
\end{equation}
is the inverse of $A$. 
 
\vspace{5pt} 
\noindent
Then
\begin{equation}
L^n = AD^n A^{-1}
\end{equation}
and we have
\begin{equation}
D^n = \left(
\begin{array}{cc}
\eta_1^n & \\ & \eta_2^n
\end{array}
\right) =
\left(
\begin{array}{cc}
\rme^{nk} & \\ & \rme^{-nk}
\end{array}
\right)
\end{equation} 
So
\begin{equation}
L^n = \frac{1}{\rme^{2k}-1}
\left(
\begin{array}{cc}
\rme^k &    1         \\
 1  &   \rme^k
\end{array}
\right) \left(
\begin{array}{cc}
\rme^{nk} & \\ & \rme^{-nk}
\end{array}
\right)
\left(
\begin{array}{cc}
\rme^k  &  -1       \\
 -1  &  \rme^k 
\end{array}
\right)
\end{equation} 
or
\begin{equation}
L^n = \left(
\begin{array}{cc}
\frac{\sinh (n+1)k}{\sinh k}  &  -\frac{\sinh nk}{\sinh k} \\
\frac{\sinh nk}{\sinh k}      &  -\frac{\sinh(n-1)k}{\sinh k}
\end{array}
\right)
\end{equation}

\end{document}